\documentclass[letterpaper,twocolumn,english,aps,prl,superscriptaddress,showpacs]{revtex4-1}
\usepackage{amsmath}
\usepackage{amssymb}
\usepackage[dvipdfm]{graphicx}
\usepackage{dcolumn}
\usepackage{setspace}

\begin{document}

\title{Excitonic effects in the optical properties of SiC sheet and nanotubes }

\author{H. C. Hsueh}

\email{hchsueh@mail.tku.edu.tw}

\affiliation{Department of Physics, Tamkang University, Tamsui, Taipei 25137,
Taiwan }

\author{G. Y. Guo}

\email{gyguo@phys.ntu.edu.tw}

\affiliation{Graduate Institute of Applied Physics, National Chengchi University, Taipei 11605, Taiwan}

\affiliation{Department of Physics, National Taiwan University, Taipei 10617, Taiwan}

\author{S. G. Louie}

\affiliation{Department of Physics, University of California at Berkeley, Berkeley,
California 94720, USA}

\affiliation{Materials Sciences Division, Lawrence Berkeley National Laboratory,
Berkeley, California 94720, USA}

\date{\today}
\begin{abstract}
The quasiparticle band structure and optical properties of single-walled
zigzag and armchair SiC nanotubes (SiC-NTs) as well as single
SiC sheet are investigated by {\it ab initio} many-body calculations using
the GW and the GW plus Bethe-Salpeter equation (GW+BSE) approaches, respectively.
Significant GW quasiparticle corrections of more than 1.0 eV to the Kohn-Sham band gaps from
the local density approximation (LDA) calculations are found. 
The GW self-energy corrections transform the SiC sheet from a indirect LDA band gap 
to a direct band gap material.
Furthermore, the quasiparticle band gaps of SiC-NTs with different chiralities
behave very differently as a function of tube diameter, and this can be attributed to 
the difference in the curvature-induced orbital rehybridization between the
different chiral nanotubes.
The calculated optical absorption spectra are dominated by discrete exciton peaks
due to exciton states with large binding energy up to 2.0 eV in the SiC sheet and SiC-NTs.
The formation of strongly bound excitons is attributed to the enhanced electron-hole interaction in these low dimensional systems.
Remarkably, the excited electron amplitude of the exciton wavefunction
is found to peak on the Si atoms near the hole position (which is on the C site) in the zigzag SiC-NTs,
indicating a charge transfer from an anion (hole) to its neighboring cations by 
photoexcitation. In contrast, this pronounced peak structure disappear in the exciton wavefunction
in the armchair SiC-NTs.
Furthermore, in the armchair SiC-NTs, the bound exciton wavefunctions
are more localized and also strongly cylindrically asymmetric.
The large excitation energy of $\sim 3.0$ eV of the first bright exciton with no dark exciton below it,
suggests that the small-radius armchair SiC-NTs be useful for optical devices working in the UV regime.
On the other hand, the zigzag SiC-NTs have many dark excitons below the first bright exciton 
and hence may have potential applications in tunable optoelectric devices ranging
from infrared to UV frequencies by external perturbations.
\end{abstract}


\maketitle

\section{Introduction}

Silicon carbide (SiC) crystallizes in either a cubic or a hexagonal form, and
exbihits interesting polytypism~\cite{mun82,wyc63}. The polytypes are made of 
identical hexagonal layers with different stacking sequences. These polytypes are 
semiconductors with a range of band gaps, from 2.39 eV in the zincblende polytype (3C) to 3.33 eV 
in the wurtzite polytype (2H)~\cite{mun82,iva92}. Furthermore, 3C and 6H SiC are used 
for high temperature, high-power and high-frequency devices~\cite{C,Wang,Rurali,Park} 
due to their unique properties~\cite{SiC}, while 6H SiC with a band gap of 2.86 eV 
is a useful material for blue light-emitting diode applications~\cite{iva92}.

Recently, SiC nanotubes (NTs) were synthesized by reaction of carbon nanotubes 
(CNTs) with SiO at various temperatures~\cite{SiCNT}.
Like other tubular materials such as C~\cite{Iijima,Zone-folding,Guo04}, BN~\cite{BNNT_TB,gyguo_BN}, 
AlN~\cite{Mei},  and GaN~\cite{Lee} nanotubes that have been synthesized previously, 
SiC-NTs display very interesting properties distinctly different from their bulks. 
Therefore, the successful synthesis of the SiC-NTs has stimulated a number of  
theoretical and experimental investigations on tubular form of SiC 
(see \cite{gyguo1,gyguo2} and references therein).
In particular, based on density-functional calculations, Miyamoto and Yu~\cite{Miyamoto}
predicted that the strain energies of SiC-NTs are lower than that of CNTs, 
and that the band gaps of SiC-NTs can be direct or indirect, depending on the chirality.
Using both tight-binding molecular dynamics and $\textit{ab initio}$ methods, 
Menon and coworkers~\cite{Menon} showed that single-walled SiC-NTs are very stable
with a large band gap.  

Unlike CNTs, SiC-NTs are polar materials and therefore, may exhibit unusual physical 
properties that CNTs may not have. For example, like BN-NTs (see, e.g., \cite{Guo_SHG,Guo07}), 
zigzag SiC-NTs may become piezoelectric, and also show second-order non-linear optical response.
A knowledge of the optical properties of SiC-NTs is important for their optical and electro-optical 
applications. Therefore, Wu and Guo have recently carried out a series of {\it ab initio} calculations 
within the density functional theory (DFT) with the local density approximation (LDA) 
in order to analyze the linear optical features and underlying band structure as well as the 
second-harmonic generation and linear electro-optical coefficients of all three types of the 
SiC-NTs.\cite{gyguo1,gyguo2}
In particular, Wu and Guo found that all the SiC nanotubes are semiconductors with exceptions of
the ultrasmall (3,0) and (4,0) zigzag tubes which may be regarded as the
thinnest conducting SiC nanowires.\cite{gyguo1} Interestingly, the band gap
of the zigzag SiC-NTs may be reduced from the energy gap of the SiC sheet
all the way down to zero by reducing the diameter, though the band gap
for all the SiC nanotubes with a diameter larger than
$\sim$20 \AA$ $ approaches that of the SiC sheet. Furthermore, DFT indicates that all the semiconducting
{\it zigzag} SiC-NTs have a direct band gap. All these suggest that they may
have interesting applications in optical and optoelectronic devices.
Nonetheless, both the armchair and chiral SiC-NTs have an indirect band gap at the DFT level.

These previous theoretical studies of the electronic and optical properties
of SiC-NTs were based on the independent-particle approximation within the DFT framework\cite{gyguo1,gyguo2}
or on semi-empirical calculations\cite{Menon}. However, many-body interactions are known to play 
an important role in low-dimensional systems such as nanotubes\cite{CNT_BSE,BNNT_BSE}, due to reduced charge
screening and enhanced electron-electron correlation. Therefore, motivated by the prospects of
optoelectronic device applications of SiC-NTs and also this theoretical issue, we have performed
{\it ab initio} calculations to study the many-body effects on the quasiparticle band gaps and optical spectra of SiC-NTs.
Indeed, we find pronounced GW quasiparticle corrections of more than 1.0 eV to the LDA band gaps
in the SiC-NTs. Further, the calculated optical absorption spectra from the SiC-NTs are dominated 
by the discrete exciton peaks due to strongly bound excitons with a large binding energy of up to 2.0 eV.
 
SiC-NTs can be considered as an atomic layer of
SiC sheet rolled up into a cylinder, and similar to the CNTs the structure of a SiC-NT is specified by the chiral
vector which is given in term of a pair of integers ($n,m$).\cite{Zone-folding}  Understanding
the optical properties of an isolated SiC sheet would help understand the many-body effects on the
optical responses of SiC-NTs. Therefore, the single atomic layer hexagonal SiC sheet is first considered here.
Furthermore, the optical properties of the two-dimensional (2D) single SiC
sheet are interesting on their own account, and, in particular, a single SiC sheet bears many similarities
to graphene which exhibits many fascinating properties.\cite{nov04,net09} 

This paper is organized as follows. In the next section, the theoretical methods and
computational details, including the ground-state, quasiparticle and electron-hole
interaction calculations, are described. In Sec. III, the calculated quasipartice band structure
and optical absorption spectrum of the isolated SiC sheet are presented.
In Sec. IV, the calculated quasiparticle band structures of the SiC-NTs are reported
and the effects of curvature and chirality on the band gaps are analyzed.
In Sec. V, the calculated optical absorption spectra of the SiC-NTs are
presented, and the excitonic effects are discussed.
Finally, the conclusions drawn from this work are given in Sec. VI.

\section{Methodology}

In the present paper, we adopt the approach of Rohlfing and Louie\cite{Rohlfing,BSE_prb} and
hence calculate the quasiparticle energies, electron-hole excitations and optical spectra from first-principles
in three steps. First, we calculate the electronic ground state 
with DFT in the LDA.\cite{Kohn} Second, we obtain the quasiparticle energies ($E^{QP}$) 
within the GW approximation for the electron self energy $\Sigma$.\cite{hybertsen98,hybertsen}
Finally, we evaluate the coupled electron-hole excitation energies and optical
spectra by solving the Bethe-Salpeter equation (BSE) of the two-particle Green's 
function.\cite{Rohlfing,BSE_prb} Details of our computations are described below.

\subsection{Ground-state properties}

An appropriate mean-field description of the ground-state properties of solids is essential
to perform quasiparticle calculations within many-body
perturbation theory. For the conventional covalent and ionic bonded materials, it is shown \cite{hybertsen} 
that eigen wavefunctions and eigen energies of the Kohn-Sham equation\cite{Kohn}in the 
LDA provide a good starting point for the many-body perturbation calculations such as within the GW approximations.
Therefore, in this paper, the ground-state electronic structure of the single SiC
sheet and SiC-NTs are first calculated by means of {\it ab initio} 
plane-wave pseudopotential method\cite{PARATEC}.
The electronic configurations of Si($3s^2,3p^2$) and C($2s^2,2p^2$)
are treated as active valence states when generating pseudopotentials
in the Kleinman-Bylander form\cite{KB} for all the calculations. The
cutoff energy of 40 Ry for the plane wave basis set is
used. Also to ensure the convergency, the
Brillouin zone (BZ) integrals for the calculations of the SiC sheet and
SiC-NTs are carried out using the 10$\times$10$\times$1 and 1$\times$1$\times$32
Monkhorst-Pack $k$-grid sampling\cite{MP}, respectively. The supercell approach
is used here such that an isolated SiC sheet (nanotube) 
is approximated by a single SiC layer (nanotube) surrounded by vacuum in a supercell. 
Sufficient intersheet and intertube separations (more than 10 $\textrm{\AA}$) are
used to prevent intersheet and intertube interactions. 
The underlying atomic positions and lattice constants are taken from Ref. \cite{gyguo1}
where the theoretical structures were obtained by the conjugate gradient method
with the atomic forces and the stress calculated from first-principles.

\subsection{Quasiparticle calculations}

Based on the ground-state Kohn-Sham wavefunctions and corresponding
eigenvalues ($E^{KS}$) calculated above, the many-body effects on 
the quasiparticle band structure characterized by the self energy 
($\Sigma$) can be evaluated by solving the Dyson equation \cite{GW,hybertsen98,hybertsen} 
\begin{equation}
\left[H_{0}+\Sigma(E_{n\bf{k}}^{QP})\right]|n{\bf k}>=E_{n\bf{k}}^{QP}|n{\bf k}>
\end{equation}
where $H_{0}$ is the Hamiltonian in the Hatree approximation, and $|n{\bf k}>$ represents the
quasiparticle wavefunction of energy $E_{n\bf{k}}^{QP}$ within one-partcile Green's function method.
In the $G^{0}W^{0}$
approximation, the vertex correction is approximated by a delta function
and the irreducible polarizibility $P^{0}$ is a convolution
of the mean field Green's function $G^{0}$, i.e., $P^{0}=-iG^{0}G^{0}$.
This gives rise to a dynamical dielectric matrix within the random phase
approximation (RPA) scheme and a generalized plasmon-pole (GPP) model\cite{hybertsen}.
Also, the screened Coulomb potential $W^{0}$ can
be obtained from the bare Coulomb interaction $v$ through the equation
$W^{0}=\upsilon/\left[1-\upsilon P^{0}\right]$. Finally, the quasiparticle
self-energy operator is given by $\Sigma= iG^{0}W^{0}$. Although
alternatives have been proposed recently to calculate the self energy going beyond this approximation, 
the $G^{0}W^{0}$ approach  is still the most efficient scheme for providing 
a successful description of the electronic
excitation and even transport properties in various semiconductors,
metals, surfaces, and novel nanomaterials\cite{G0W0_solid,G0W0_nano,GW_transport}.
Therefore, in this paper, we adopt the $G^{0}W^{0}$ approach to calculate the
quasiparticle properties of the single SiC sheet and SiC-NTs.
Here a rather dense $k$-point sampling for the SiC sheet (18$\times$18$\times$1)
and also for the SiC-NTs (1$\times$1$\times$32) is used to ensure
the calculated quasiparticle energies converged to within $0.05$ eV. 

\begin{figure}
\includegraphics[angle=270,scale=0.4]{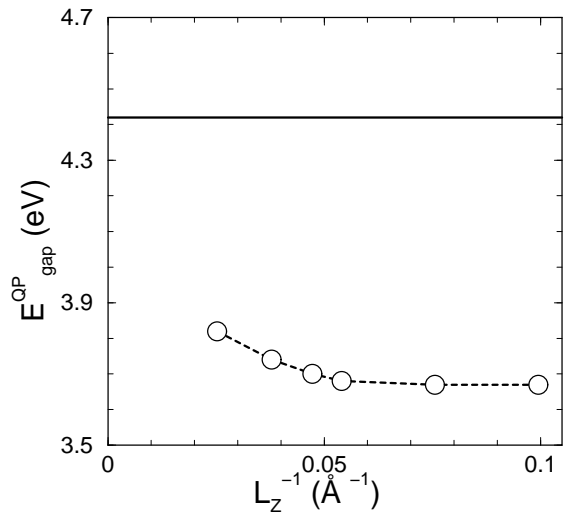}
\caption{\label{fig:fig1} The calculated quasiparticle band gap of
a SiC sheet (open symbols)
as a function of the inverse of intersheet separation ($L_{z}$). 
The converged band gap indicated as the solid line is derived
by employing the Coulomb truncation scheme. The dashed line linking
the open circles is a guide to the eye only.}
\end{figure}

\begin{table}[b]
\caption{The band gap ($E_{gap}$) and quasiparticle energy differences between the 
conduction band minimum ($c$) and valence band maximum ($v$) at some high symmetry $k$-points of the SiC sheet
from the LDA and GW [without/with the Coulomb truncation (no trunc.)/(trunc.)] calculations. All energies are in eV.}
\begin{tabular}{cccc}
\hline
\hline 
 & LDA & \multicolumn{2}{c}{GW} \\
 & ----------- & \multicolumn{2}{c}{------------------------} \\
 & (no trunc.) & (no trunc.) & (trunc.) \\ \hline
$E_{gap}$ & 2.51 & 3.66 & 4.42 \\
$K^{v}\rightarrow K^{c}$ & 2.56 & 3.66 & 4.42 \\
$\Gamma^{v}\rightarrow\Gamma^{c}$ & 4.46 & 5.58 & 6.04 \\
$M^{v}\rightarrow M^{c}$ & 3.23 & 4.60 & 5.40 \\
\hline
\end{tabular}
\end{table}

A proper description of the dynamical screening properties of a solid 
plays an important role in obtaining reliably its quasiparticle properties. 
In low dimensional systems, such requirement becomes crucial 
due to the highly anisotropic screening interactions associated mainly with the
vacuum region between the isolated units in the supercell scheme.
A larger vacuum region is preferred to reduce the residual unphysical
inter-unit interaction and eventually ends up with an ideal vacuum screening of the isolated units.
For example, the band gap of an isolated SiC sheet, as shown in Fig. 1,
would increase as the screening from neighboring layers is reduced by increasing the 
separation ($L_{z}$) between the SiC layers in the neighboring supercells.
Furthermore, the convergence of the band gap with respect to the spatial 
separation of the neighboring SiC layers is very slow. Due to the computation
constraints, it is impossible to obtain the fully converged band gap 
simply by expanding the inter-layer separation. Therefore,
an efficient truncation scheme\cite{truncation_SGL,truncation_SIB} to the Coulomb interaction in which 
a step function at the boundary of the supercell is introduced, will be adopted
in the present GW and BSE calculations to remove the long-range Coulomb 
interaction between the structure and its images.
Fig. 1 clearly demonstrates that the truncation scheme indeed offers
an efficient means to achieve the converged quasiparticle
band gap for the low-dimensional structures. Table I further shows that the
quasiparticle band gap and also quasiparticle energy differences from the GW calculations with or without
the Coulomb truncation scheme can differ significantly. 

\subsection{Electron-hole excitations}

The optical properties of a solid, associated with the
interaction between light and the electronic excitations of a system, 
are described by the frequency-dependent macroscopic dielectric function
$\epsilon\left(\omega\right)$. For example, the optical
absorption spectrum is determined by the imaginary part of the dielectric
function, $\epsilon_{2}\left(\omega\right)$. According to the Fermi
Golden rule, the transition energy associated with  a peak in the optical absorption 
spectrum can be estimated from the energy difference between the associated
electron and hole states. The corresponding oscillator strength can be
obtained from the optical transition matrix elements derived from
the electronic ground and excited states involved. Since the incident
photons carry negligible momentum, the peaks in the optical absorption spectrum 
can be considered as direct interband transitions between the 
occupied and unoccupied electronic states without any crystal momentum transfer.
Therefore, if ignoring the electron-hole interaction, both linear and
nonlinear optical spectra can be calculated in an independent-particle
approximation by including only the vertical interband transitions between
the Kohn-Sham states\cite{gyguo1,gyguo2}. In the GW+RPA approach, 
the dielectric function calculated based on the quasiparticle energies as described
above, gives the interband optical absorption spectrum beyond the LDA, but still 
without electron-hole interaction (or excitonic) effects. 

However, for the systems with strong excitonic effects, the optical
properties may be dominated by excitons which are composed
of strongly correlated electron-hole pairs of the systems. The connection of the exciton
energies $\Omega^{S}$ and corresponding electron-hole amplitudes $A_{vc{\bf k}}^{S}$
of the correlated electron-hole excitations $S$ is governed by the Bethe-Salpeter
equation\cite{BSE_prb}
\begin{equation}
\left(E_{c{\bf k}}-E_{v{\bf k}}\right)A_{vc{\bf k}}^{S}+\underset{v'c'{\bf k}'}
{\sum}<vc{\bf k}|K^{ph}|v'c'{\bf k}'>A_{vc{\bf k}}^{S} =\Omega^{S}A_{vc{\bf k}}^{S}
\end{equation}
where $E_{v{\bf k}}$($E_{c{\bf k}}$) denotes the quasiparticle (e.g., GW) eigenvalues 
of valence (conduction) bands at a specific ${\bf k}$-point and $K^{eh}$ is the 
kernel describing the interaction between excited electrons
and holes. Therefore, a more realistic two-particle picture of the optical excitations
including excitonic effects in these systems should be provided by means of
the GW+BSE method. To obtain the converged optical spectra, 
in the present work, the kernel $K^{eh}$ of the SiC sheet (SiC-NTs) is computed on a 
sparse $k$-point grid of 18$\times$18$\times$1 (1$\times$1$\times$32) and then interpolated\cite{Rohlfing}
onto a denser $k$-point grid of 36$\times$36$\times$1 (1$\times$1$\times$64).

Instructively, the electron-hole amplitude in real space can be expanded in the quasielectron
and quasihole basis $\left\{ \phi_{c{\bf k}}({\bf r}_e),\phi_{v{\bf k}}({\bf r}_h)\right\}$,
\begin{equation}
\Phi_{S}({\bf r}_e,{\bf r}_h)=\underset{\bf k}{\sum}\underset{c,v}
{\sum}A_{vc{\bf k}}^{S}\phi_{c{\bf k}}({\bf r}_e)\phi_{v{\bf k}}({\bf r}_h)
\end{equation}
and the corresponding exciton states can be visualized in real space. Because of the complexity of the six-dimensional 
exciton wavefunction, a simpler distribution of the electron amplitude square with the hole 
position fixed, i.e., $|\Phi_{S}({\bf r}_e,{\bf r}_h=0)|^{2}$, is usually used to reveal 
the essence of the electron-hole correlation of the exciton.\cite{CNT_BSE,BNNT_BSE}

Finally, as noticed before\cite{Guo04,gyguo1}, the employment of a 3D supercell method
for reduced-dimensional systems will generate an arbitrary volume effect
on the dielectric function computation. 
In order to resolve this ambiguity and compare directly to experimental measurements, we calculate the imaginary part of
polarizability per unit area (in units of nm) 
for the SiC sheet, $\alpha_{2}\left(\omega\right)$, which is derived from 
the calculated dielectric susceptibility, $\chi=\left(\epsilon-1\right)/4\pi$, multiplied by the distance 
between two neighboring SiC sheets.\cite{Yang09} 
Moreover, for the one-dimensional (1D) nanotubes, the measured optical spectrum 
is determined by the imaginary part of the optical polarizability per
single tube, $\alpha_{2}\left(\omega\right)$. This $\alpha_{2}$ is equal to the $\epsilon_{2}^{calc.}$
multiplied by the cross-sectional area of the supercell perpendicular to the tubular 
axis ($\Lambda_{sc}^{\perp}$), 
i.e., $\alpha_{2}=\Lambda_{sc}^{\perp}\left(\epsilon_{2}^{calc}\right)/4\pi$.


\section{Electronic and optical properties of SiC sheet}

In the recent DFT-LDA calculations, a stable non-buckled
graphene-like monolayer has been found for the isolated SiC
layer\cite{gyguo1,Sahin}. 
Despite the structural similarity, the unique massless Dirac fermion
feature in graphene is no longer present in the SiC sheet because of
the different ionicities of the Si and C atoms. This heteropolar
character induces an energy gap of $2.56$ eV (the LDA value) at the $K$ point of the
hexagonal Brillouin zone, as shown in Fig. 2. 
The quasiparticle corrections to the 2D SiC sheet state are obtained
through the GW calculations described above.
The calculated quasiparticle band gaps and energy differenes 
at some high symmetry k-points are summarized in Table I.
It is clear from Table I that the GW corrections to the LDA
eigenvalues are rather significant, being more than 1.0 eV.
We note that the GW correction to the LDA band gap of bulk 2H SiC is 
around 1.0 eV.\cite{2H-SiC_gap,Hsueh} 
Strikingly, the GW corrections transform the indirect LDA band gap ($K\rightarrow M$) 
into a direct band gap ($K\rightarrow K$) for the SiC sheet. 
The nature of the direct band gap in the 2D SiC sheet indicate 
that the lowest energy exciton in the SiC sheet is optically active (bright). 
Furthermore, based on the simple zone-folding approximation, we could expect no 
momentum-mismatch induced optically inactive (dark) excitons below the lowest 
energy bright exciton in the SiC-NTs\cite{GaN_Sohrab}. This anticipation i
is indeed confirmed by our GW+BSE calculations 
for both the SiC sheet and armchair SiC-NTs presented below. However, our GW+BSE calculations 
also show that this is not the case for very small zigzag SiC-NTs, as will be discussed in Sec. V. 

Figure 2 shows the LDA band structure (left panel) and also GW corrections
to the LDA eigenvalues (right panel) of the 2D SiC sheet.
We note that the optical dielectric function of the 2D SiC sheet is highly
anisotropic (see, e.g., Ref. \onlinecite{gyguo1}). In particular, there is
only weak dynamical screening for the electric field perpendicular to
the sheet in the energy range below 6 eV, whereas the dynamical screening for 
the in-plane electric field is strong in this energy range
(see Fig. 1 in Ref. \onlinecite{gyguo1}).
Therefore, many-body interaction effects in the  $\pi$ band (red line in Fig. 2) associated with the $p_z$ orbitals,
which extend into the vacuum region from the sheet, would be less
screened. Consequently, the  $\pi$ band would have a larger quasiparticle 
correction and hence have a larger difference between the quasiparticle energies 
and LDA eigenvalues ($E^{QP}-E^{LDA}$) (red circles in Fig. 2).
On the other hand, because the in-plane $\sigma$ bonds are mainly confined to
the 2D SiC sheet, electric screening effects on the $\sigma$ bands (blue lines in Fig. 2) 
are more significant, and hence GW corrections are smaller (blue diamonds in Fig. 2). 
Furthermore, due to weak many-body interaction in nearly free electron (NFE) states 
which mainly lie more than a few $\text{\AA}$s above and below the SiC sheet,
the GW corrections to these conduction bands are small, being less than $0.3$ eV
(green squares in Fig. 2). This is similar to the NFE states in graphene\cite{Li-Yang}. 
These sophisticated screening effects enforce the notion that the quasiparticle 
calculations rather than the simple scissor-operator approximation are
necessary for low-dimensional systems such as the 2D SiC sheet.

\begin{figure}
\includegraphics[angle=270,scale=0.40]{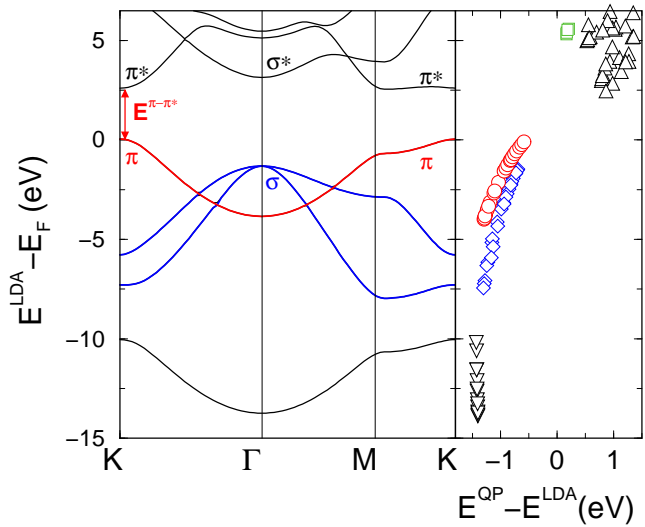}
\caption{(Color online) The LDA band structure (left panel) and GW corrections to
the LDA eigenvalues (right panel) of the graphitic SiC sheet. The $\pi$ and $\sigma$ bands 
(left panel) are denoted as red and blue lines, respectively. In the right panel,
the GW corrections for the $\pi$ and $\sigma$ bands are represented by red circles 
and blue diamonds, respectively, whereas the GW corrections of the 
NFE states are represented by green squares. The LDA $E^{\pi-\pi*}$ gap is marked 
as an arrow in the left panel. The valence band maximum is aligned at 0 eV.}
\end{figure}

The excitonic effects on the optical properties of the SiC sheet can be examined
by comparing the GW+RPA results with the GW+BSE calculations. Because of the huge 
depolarization effect in the 2-D planar geometry for light polarization perpendicular
to the plane, we will focus on the optical absorption spectrum for light polarization 
parallel to the plane. Figure 3 shows the optical spectra from both the GW+RPA and
GW+BSE calculations (a) as well as the $E_1^{\pi-\pi*}$ exciton wavefunction 
(b) of the SiC sheet. 
In Fig. 3(a), the first prominent peak located at $E_{1}^{\pi-\pi*}=3.25$ eV 
in the GW+BSE absorption spectrum comes from a bright 
exciton due to the excitation between the $\pi$ and $\pi^{*}$ states 
at the $K$ point [see Fig. 2 (left panel)]. This is a strongly bound exciton
with a large binding energy of $E_{1}^{Bind}=1.17$ eV as measured by the difference
between the $E_{1}^{\pi-\pi*}$ and the onset energy of the GW corrected electron-hole
continuum ($E^{\pi-\pi*}=4.42$ eV). We note that the theoretical exciton binding energy 
in bulk 2H SiC is only 0.1 eV.\cite{Hsueh} This shows clearly that the reduced 
dimensionality of a SiC sheet strongly confine the quasiparticles, and this significantly
enhance the overlap between the electron and hole wave functions and hence the
electron-hole interaction. In addition to this quantum confinement,
the presence of the vacuum region reduces the screening and hence provides extra 
contribution to the large excitonic effect in the SiC sheet. 
An analysis of $\int |\Phi_{S}({\bf r}_e,{\bf r}_h)|^{2}d{\bf r}_e$ and  
$\int |\Phi_{S}({\bf r}_e,{\bf r}_h)|^{2}d{\bf r}_h$ shows that the hole is 
mainly resided in the C-sublattice whereas the excited electron is on the Si-sublattice.
In order to elucidate
the extent of the $E_{1}^{\pi-\pi*}$ exciton wavefunction, we plot the electron
amplitude square with the hole position fixed at the position slightly above
the C atom (see Fig. 3b). Clearly, the electron orbital is of the $p_{z}$ character,
and distributes mainly on the nearest neighbor Si atoms around the
hole to form a fairly localized exciton. 

\begin{figure}
\includegraphics[width=8cm]{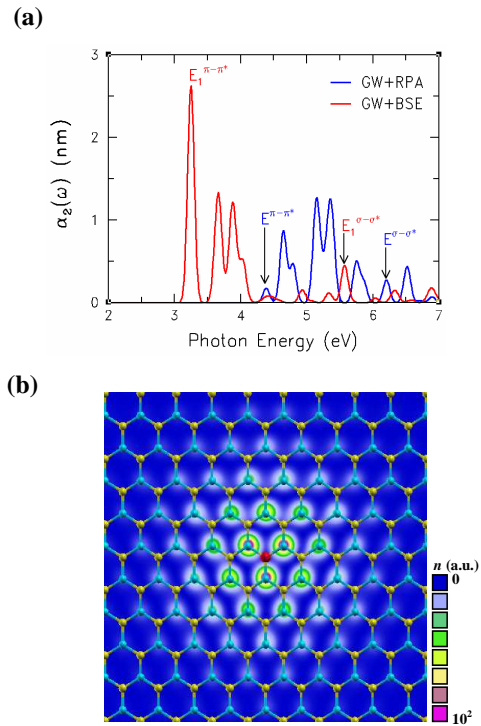}
\caption{(Color online) (a) The optical polarizability per unit area  from both  GW+RPA (blue)and GW+BSE (red) 
calculations and (b) the $E_{1}^{\pi-\pi*}$ exciton 
wavefunction of the SiC sheet. In (a), the theoretical spectra are broadened 
with a Gaussian smearing width of 0.15 eV. In (b), the hole position 
(red sphere) is fixed at the top of a C atom (yellow sphere) and the squared electron 
amplitude ($n =|\Phi_{S}({\bf r}_e,{\bf r}_h=0)|^{2}$ in an arbitrary unit (a.u.)) 
is mainly distributed on the Si atoms (cyan spheres) next to the hole.}

\end{figure}

In the optical absorption spectrum of the SiC sheet $\alpha_{2}(\omega)$ from the GW+BSE calculations,
the low-energy features in the energy range from 3.0 to 5.0 eV (Fig. 3a) are
dominated by the $\pi\rightarrow\pi^{*}$ transitions near the $K$ point
at the zone edge (Fig. 2), whereas the pronounced absorption peak at $5.83$ eV is 
mainly due to the $\sigma\rightarrow\sigma^{*}$ transition at the zone center.
As shown in Fig. 3(a), the electron-hole interaction not only triggers
a red-shift of the onset optical transition energies
but also modifies their relative absorption intensities. 
Furthermore, because the GW corrections transform the indirect LDA band gap of
the SiC sheet into the direct band gap, as mentioned above, no dark exciton 
was found in the optical spectrum until the photon energy up to $4.2$ eV 
which is above the lowest bright exciton ($E_{1}^{\pi-\pi*}$).

\section{Quasiparticle band structure of SiC nanotubes}

\begin{figure}
\includegraphics[angle=270,scale=0.4]{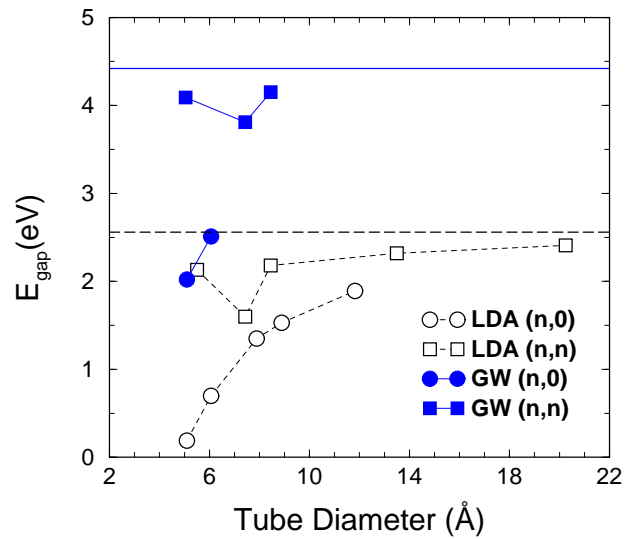}
\caption{(Color online) The fundamental band gaps of some zigzag ($n$,0)
and armchair ($n$,$n$) SiC nanotubes as a function of the tube diameter
from the GW (solid symbols) and LDA\cite{gyguo1}(open symbols)
calculations. The band gap of the isolated SiC sheet obtained from 
the GW and LDA calculations is also displayed as the solid and dashed line, 
respectively. Dotted lines between the symbols are a guide to the eye only.}
\end{figure}

An intuitive approach to construct a nanotube with a specific chirality
is simply to roll up a layer of the graphitic sheet along a specific
lattice vector into a nanocylinder. This geometrical connection
between a single-wall nanotube and the isolated sheet has inspired the
application of the zone-folding method to construct the zeroth-order electronic band structure
and phonon dispersions for single-wall nanotubes\cite{Zone-folding}.
However, the zone-folding scheme alone cannot address the significant
orbital rehybridization due to the curvature effect and also the 
many-body interaction effects 
in the nanotubes mentioned above.
To include these effects, we have calculated the diameter-dependence 
of the fundamental quasiparticle band gap of some small-radius zigzag and
armchair SiC-NTs, as compiled in Fig. 4. First of all, we note that
all the SiC-NTs are semiconducters, and this may be attributed
to the different potentials of the Si and C atoms, being similar to the
case of the BN nanotubes\cite{BNNT_TB}. Secondly, an asymptotic
convergence of the calculated band gaps of the zigzag and armchair SiC-NTs
as a function of tube size towards that of the isolated SiC sheet is observed in
both the LDA and GW calculations. 

\begin{figure}
\includegraphics[width=8cm]{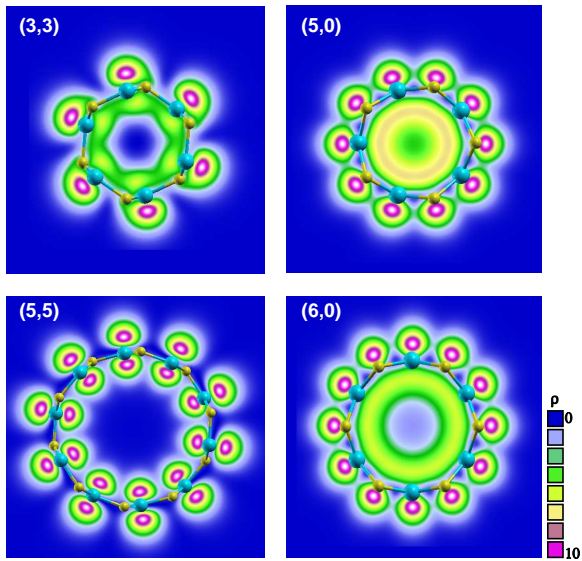}
\caption{(Color online) The charge density contours ($\rho$ in $10^{-2}{\AA}^{-3}$)of the states at the conduction band
minimum (CBM) at the 1D Brillouin zone boundary ($Z$ point) of the (3,3) and (5,5) SiC-NTs 
and also at the zone center ($\Gamma$ point) of the (5,0) and (6,0) SiC-NTs. The 
Si and C atoms are represented by the cyan and yellow spheres, respectively.}
\end{figure}

Interestingly, in contrast to the reciprocal radius dependence of the band gap 
in CNTs, we find a peculiar band gap reduction with the decrease of tube size 
for both zigzag and armchair SiC-NTs. This band gap reduction 
of SiC-NTs with the increasing tubular curvature can be attributed to
the curvature-induced hybridization between the $\pi^{*}$ and $\sigma^{*}$ orbitals
near the conduction band minimum (CBM) in the small radius SiC-NTs, 
similar to that found in the very small diameter CNTs.\cite{Blase}
To illustrate this point, we display the charge density contours of the CBM
of the zigzag ($n$,0) and armchair ($n$,$n$) nanotubes at the 1D Brillouin zone 
center ($\Gamma$ point) and zone boundary ($Z$ point), respectively, in Fig. 5. 
Figure 5 shows that in the small  zigzag ($n$,0) SiC-NTs such as (5,0) and (6,0) 
nanotubes, a strong mixture of the $\pi^{*}$ and $\sigma^{*}$ states
in the SiC sheet now exists inside the tube to form a ring-like charge distribution,
whereas the feature of the $p_{z}$ state is floating above the Si and C atoms 
outside the SiC-NT even under considerable structural distortion.
Therefore, a substantial band gap reduction upon increasing the tube curvature
can be expected. Indeed, this is confirmed by our LDA and quasiparticle
calculations (see circle symbols in Fig. 4). In contrast, a much smaller
degree of the orbital hybridization in the armchair SiC-NTs is induced by
the folding-curvature, as revealed by the cylindrically symmetric distribution 
of the Si $p_{z}$ component of the CBM at the Brillouin zone boundary ($Z$ point) 
in the (5,5) nanotube (Fig. 5). This gives rise to a minor size dependence
of the band gap (square symbols in Fig. 4) for the armchair SiC-NTs. 
Nevertheless, a spectacular curvature effect in terms of strongly enhanced 
$\pi^{*}-\sigma^{*}$ hybridization does occur in the ultrasmall armchair (3,3) SiC-NT
(see Fig. 5). This unusual orbital rehybridization effects on the band gap have 
also been recognized before in other ionic group III-V nanotubes\cite{gyguo_BN,GaN_Sohrab}. 

The quasiparticle corrections to the Kohn-Sham eigenvalues
from the GW calculations for two ultrasmall (radius being $\sim5$ \AA{}) armchair
(3,3) and zigzag (5,0) SiC-NTs are given in Fig. 6, together with their LDA band structures.
In the (3,3) SiC-NT, the GW corrections widens the LDA indirect band gap ($2.13$ eV)
to the direct one of $4.09$ eV, and also modify the band dispersions signicantly 
as a result of the complicated energy dependence of the GW corrections 
(see open circles in Fig. 6). Meanwhile, the small direct band gap (0.19 eV) of the (5,0) SiC-NT predicted
by the LDA is increased to $2.02$ eV in the GW calculations. Furthermore, the GW corrections
depend rather sensitively on the electronic state character, as illustrated in Fig. 6
for the (5,0) SiC-NT (open squares). 
We note that the curvature-induced hybridization between the $\pi$ and $\sigma$ states 
in these ultrasmall SiC-NTs is responsible for the mixture
of the quasipaticle corrections of the otherwise orthogonal $\pi$ and $\sigma$
states near the top of the valence band in the SiC sheet.
This hybridization would become weaker gradually as the tube radius increases,
and eventually diminish in the large tube radius limit where the self-energy corrections 
would be the same as that of the SiC sheet (as shown Fig. 2). 
The NFE tubular states are found in both (3,3) and (5,0) tubes as well
as the small-radius chiral SiC-NTs. However, these NFE states lie in the rather high
energy region (more than 5 eV above the band gap region from the LDA calculations)
and hence play a minor role in the optical absorption spectrum. Therefore, they 
are not included in the present GW and GW+BSE calculations.
Nonetheless, the NFE states can be shifted downward 
toward the band gap regime by doping electrons in the nanotubes\cite{NFE} and
consequently may significantly affect the optical properties of the $n$-type SiC-NTs.

\begin{figure}
\includegraphics[angle=270,scale=0.4]{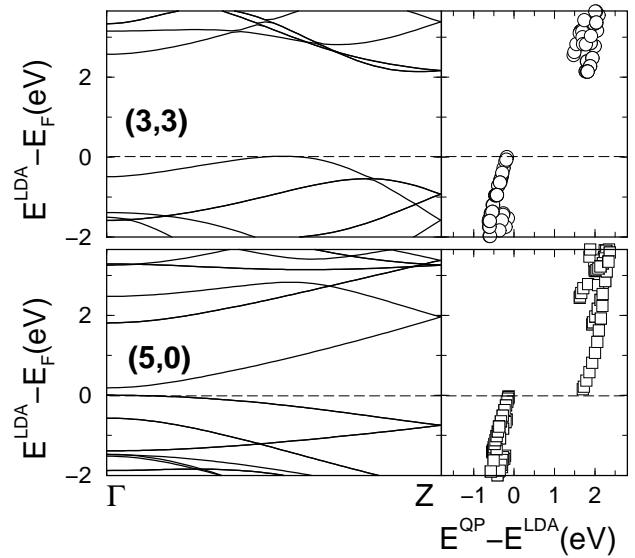}
\caption{The energy differences between the quasiparticle eigenvalues($E^{QP}$) and
LDA eigenenergies ($E^{LDA}$) of the armchair (3,3)
(upper panel) and zigzag (5,0) (lower panel) SiC-NTs. The corresponding LDA
band structures are shown in the left panels for reference.
The dashed lines denote the valence band maximum at 0 eV.}

\end{figure}

\section{Excitonic effects in SiC nanotubes}

In quasi-one dimensional nanotube systems, the wavefunctions of 
quasielectrons and quasiholes can overlap significantly because
of quantum confinement, and therefore, electron-hole interaction
may be considerably enhanced. At the same time, the screening among the quasiparticles 
can be systematically changed because of curvature  
and reduced-dimensionality effects. For example, the phenomenon of 
anti-screening was discovered recently in the CNTs.\cite{Jack} 
Therefore, the absorption features associated
with the exciton formation can become prominant in the optical spectra in SiC-NTs. 
To study the excitonic effects on the optical properties of SiC-NTs,
we have performed GW+BSE calculations by taking into account the
electron-hole interaction kernel for several SiC-NTs of different sizes
and different chiralities. 

\begin{figure}
\includegraphics[angle=0,scale=0.5]{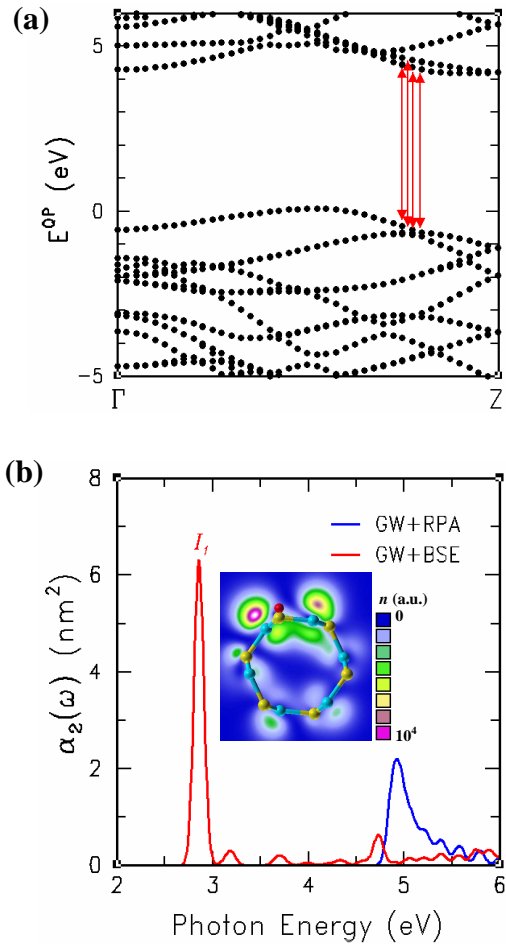}
\caption{(Color online) (a) The quasiparticle band structure and (b) 
the optical absorption spectra of the armchair (3,3) SiC-NT. 
In (b), the optical spectra from GW+RPA and GW+BSE calculations
are denoted as the blue-dashed and red-solid curves, respectively. 
The theoretical absorption spectra were broadened with a Gaussian of 0.15 eV. 
The lowest-energy bright exciton ($I_{1}$) is formed by the mixture 
of four interband transitions as indicated by the red arrows in (a). 
In the inset in (b), the squared electron amplitude ($n=|\Phi_{S}({\bf r}_e,{\bf r}_h=0)|^{2}$ in an arbitrary unit (a.u.)) of the
$I_{1}$ exciton on a cross-sectional tubular plane 
with the hole fixed at the position of the red sphere  
(see text) are displayed. Si and C atoms are represented by cyan and
yellow spheres, respectively.}
\end{figure}
 
In Fig. 7, the quasiparticle band structure and the imaginary part of 
the optical polarizability for the ultrasmall armchair (3,3) SiC-NT
are presented. Because of the strong depolarization effects in nanotubes\cite{aji94},
only the optical response for light polarized along the tube axis is significant.
Therefore, only this polarization is considered here.
For comparison, the optical absorption spectrum obtained without
the electron-hole interaction in the GW+RPA scheme is also displayed
in Fig. 7 (the blue-dashed curve in Fig. 7b).
The onset energy (4.86 eV) of this independent-particle
optical spectrum corresponds to the lowest-energy dipole-allowed 
vertical transition ($E^{I_{1}}$) from the valence to conduction
bands near 2/3 of the $\Gamma-Z$ line in the Brillouin zone
which corresponds to the $K$ point in the hexagonal BZ of the SiC 
sheet (see Fig. 7a).

\begin{table}
\caption{The binding energies ($E_{B}$) of the first bright exciton ($I_{1}$)
and lowest energy dark exciton ($K_{1}$) below $I_{1}$ for
the single-wall zigzag and armchair SiC nanotubes (NTs) considered here.
The exciton binding energy ($E_{B}$) is calculated as the
energy difference between the onset energy ($E$) of the corresponding interband
transition continuum and the excitation energy ($\Omega$). The tube diameter ($D$)
and quasiparticle energy gap ($E_{gap}^{QP}$) are listed for comparison.
Symbol (i) denotes an indirect band gap in the armchair SiC-NTs. 
Moreover, the results for the SiC sheet which can be
considered as the limiting case of an infinite diameter SiC-NT,
are also listed for comparison. All the energies are in the unit of eV.}
\begin{tabular}{ccccccccc}
 &  &  &  &  &  &  &  & \tabularnewline
\hline
\hline 
 & $D$ (\AA) & $E_{gap}^{QP}$ & $E^{K_{1}}$ & $E^{I_{1}}$ & $\Omega^{K_{1}}$ & $\Omega^{I_{1}}$ & $E_{B}^{K_{1}}$ & $E_{B}^{I_{1}}$\tabularnewline
\hline 
 NTs &  &  &  &  &  &  &  & \tabularnewline
(3,3) & 5.05 & 4.09(i) &  & 4.86 &  & 2.86 &  & 2.00\tabularnewline
(4,4) & 7.43 & 3.81(i) &  & 4.75 &  & 3.12 &  & 1.63\tabularnewline
(5,5) & 8.45 & 4.15(i) &  & 4.55 &  & 3.13 &  & 1.42\tabularnewline
(5,0) & 5.10 & 2.02 & 2.02 & 4.92 & 0.28 & 3.21 & 1.64 & 1.71\tabularnewline
(6,0) & 6.06 & 2.51 & 2.51 & 5.70 & 0.95 & 4.24 & 1.56 & 1.46\tabularnewline
 &  &  &  &  &  &  &  & \tabularnewline
Sheet & $\infty$ & 4.42 &  & 4.42 &  & 3.25 &  & 1.17\tabularnewline
\hline
\hline 
 &  &  &  &  &  &  &  & \tabularnewline
\end{tabular}
\end{table}

When the interaction between electrons and holes is included,
excitons with specific excitation energies ($\Omega$) are formed
to give rise to the prominent photoabsorption peaks. The binding energy
of the lowest-energy exciton is defined as the energy difference between 
the continuum onset and corresponding excitation energy. For the armchair
(3,3) SiC-NT, the optical absorption spectra from
the GW+BSE calculations is dominated by the discrete exciton peaks 
[see the red curves in Fig. 7(b)]. The first bright exciton ($I_{1}$)
appearing at $2.86$ eV with a large binding energy ($E_{B}^{I_{1}}$)
of $\sim2.0$ eV is formed by a mixture of four pair of interband transitions,
as indicated by the arrows in Fig. 7(a). These four different interband transitions couple strongly
to make the electron amplitude more localized and
highly asymmetric around the hole position located outside the tube but near
the C atom, as displayed in the inset
in Fig. (b), thereby resulting in a strongly bound character of exciton
$I_{1}$. At the same time, the large electron-hole overlap in the (3,3) SiC-NT
lifts the double degeneracy of the $E_{1}^{\pi-\pi*}$ exciton 
in the 2D SiC sheet to create the bright $I_{1}$ exciton and a
dark exciton with a higher (by $0.02$ eV) excitation
energy. Similar excitonic effects are also found in the larger 
armchair (5,5) SiC-NT (see Table II). However, the lowest optically active exciton
in the (5,5) SiC-NT has a smaller binding energy of $E_{B}^{I_{1}}=1.42$ eV,
because the reduced electron-hole overlap due to the decreased curvature. 
Moreover, apart from the bright bound and resonant excitons shown as red peaks in Fig 7(b), 
there also exist many dark excitons among the bright ones. However, 
we find no dark exciton below the $I_{1}$ exciton 
in the armchair SiC-NT. As mentioned before, this can be anticipated from
the fact that the isolated SiC layer has a direct band gap at the $K$-point.
In this case, there is no momentum mismatch when the SiC sheet is rolled 
up to form an armchair SiC-NT, and hence the possibility
of the formation of the optically inactive lowest-energy exciton
diminishes\cite{GaN_Sohrab}. 

\begin{figure}
\includegraphics[angle=0,scale=0.5]{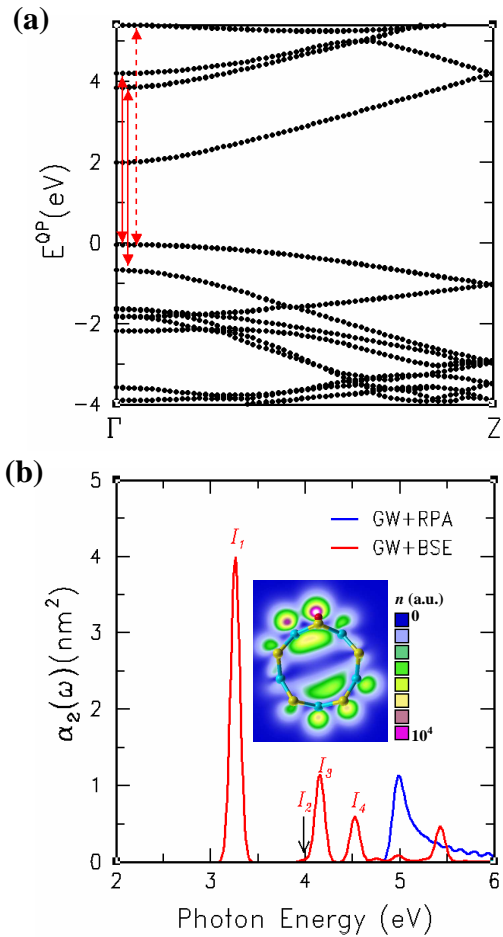}
\caption{(Color online) (a) The quasiparticle band structure and (b) optical absorption
spectrum (broadened with a Gaussian width of 0.15 eV) of the zigzag (5,0)
SiC-NT. In (b), the first bright exciton, $I_{1}$, is mainly related to
the interband transition indicated as a red-solid arrow in (a) whereas
the transition by the red-dashed arrow 
in (a) corresponds to the third ($I_{3}$) bright exciton in (b).
The inset in (b) shows the squared electron amplitude ($n$ in an arbitrary unit) of $I_{1}$ with 
the fixed hole position marked by a red sphere.
The Si and C atoms are represented by the cyan and yellow spheres, respectively.}
\end{figure}

Next, let us concentrate on the excitonic effects on the optical absorption
in the small radius zigzag SiC-NTs. In order to scrutinize the chirality
effects, we performed GW+RPA and GW+BSE calculations of the zigzag (5,0)
SiC-NT with a tube radius comparable to that of the armchair (3,3) SiC-NT.
The quasiparticle band structure and optical polarizability
$\alpha_{2}$ of the (5,0) SiC-NT is displayed in Fig. 8 (a) and (b), respectively.
In Fig. 8 (b), label $I_{j}$ denotes the $j$th optically allowed excited state.
The inset in Fig. 8 (b) is a contour plot of the electron charge density 
distribution of the $I_{1}$ exciton with respect to a fixed hole position.
Like the (3,3) SiC-NT, the photoexcited spectrum of the (5,0) SiC-NT is characterized
by the discrete exciton peaks. The first bright exciton $I_{1}$ is the most 
prominent feature in the spectrum, and consists of the different interband transitions 
near the zone-center $\Gamma$-point (represented by the solid arrow). 
As a result, the charge density distribution is not cylindrically symmetric.
Similar results are found in the (6,0) SiC-NT, and the corresponding excitation 
energies are listed in Table II. Interestingly, although the (5,0) and (3,3) SiC-NTs have
a similar curvature, the relatively smaller binding energy (0.3 eV) of the $I_{1}$ 
exciton in the (5,0) SiC-NT indicates a rather delocalized exciton. This 
pronounced chirality effect can be attributed to the considerable orbital 
rehybridization in the zigzag SiC-NTs emphaszied above. 
This orbital rehybridization induces a significant band-gap reduction giving rise to the 
increasing of the effective screening, and hence reduces the binding energy. 
Furthermore, according to the zone-folding scheme for nanotubes, the lowest energy 
optically active interband transition at the $K$ point of the 2D sheet can be folded into 
the $\Gamma$ point to form a bright exciton with the lowest excitation energy in 
zigzag nanotubes. However, in the very small-radius zigzag SiC-NTs such as the (5,0) nanotube, 
the strong curvature-induced orbital rehybridization has brought a
higher conduction band down to the GW band gap ($0.0 \sim 4.0$ eV) region 
[see Fig. 8 (a)]. Since the optical transitions from the top valence bands to this
conduction band are dipole-forbiden, some low energy dark excitons including
the $K_{1}$ dark exciton, appear in the small-radius zigzag SiC-NTs.
As shown in Table II, this low-energy 
$K_{1}$ dark exciton is also rather localized with a large binding energy, and this 
could make the zigzag SiC-NTs candidates for the tunable optical devices for
temperature or external field sensors.

Finally, we explore the extent of the exciton wavefunction localization in the different SiC-NTs
by comparing the intensity of their electron-hole wavefunctions in real space. In Fig. 9, 
the isosurface plots of the electron density with the fixed hole position 
(the red cross in the figure) of the first 
bright exciton ($I_{1}$) of the armchair SiC-NTs suggest a significant local distribution 
in the direction along the circumference of the tube. Mixing of excitations to different subband pairs in the small-radius
SiC-NTs is responsible for this tightly bound character. In addition,
a spatial localization of the exciton wavefunction along the tubular direction
is also revealed by integrating out the electron coordinates
in the perpendicular plane (red curves in Fig. 9). In contrast with
the carbon nanotubes, the deviation from the one-dimensional nature of the low-energy
exciton is mainly attributed to the strongly mixing of the interband transitions.
Similar tightly bound behavior of the lowest-energy
bright exciton has also been discovered in the BN-NTs\cite{BNNT_BSE}.
In the zigzag SiC-NTs, we find the similar anisotropic bound nature
of the first bright exciton. The slightly broadened distribution of the integrated 
wavefunction intensity (see the blue curves in Fig. 9) suggests a less bound character 
in the zigzag SiC-NTs. 
The distinct spiky features in the curve are located on the cation (Si atom) planes
and they indicate a charge transfer from an anion (hole) to its neighboring cations
by photoexcitation. On the other hand, the pronounced peak structures are not observed
in the armchair SiC-NT because of the specific atomic arrangement 
dictated by its particular chirality. Furthermore, the asymmetric distribution of
$|\Phi|^{2}$ with respect to the fixed hole position, is a consequence of
the different nearest-neighbor cation-anion distances caused by rolling
up the ideal symmetric 2D SiC sheet. This curvature-induced spatial
symmetry breaking is also responsible for the special nodal-like feature
near the hole position in the smallest armchair (3,3) SiC-NT. Nonetheless,
by comparing the distribution of the exciton wavefunction of the (3,3) SiC-NT 
with the larger (5,5) nanotube, we note that this symmetry breaking will 
decrease gradually as the curvature decreases or the tube diameter increases. 

\begin{figure}
\includegraphics[angle=0,scale=0.5]{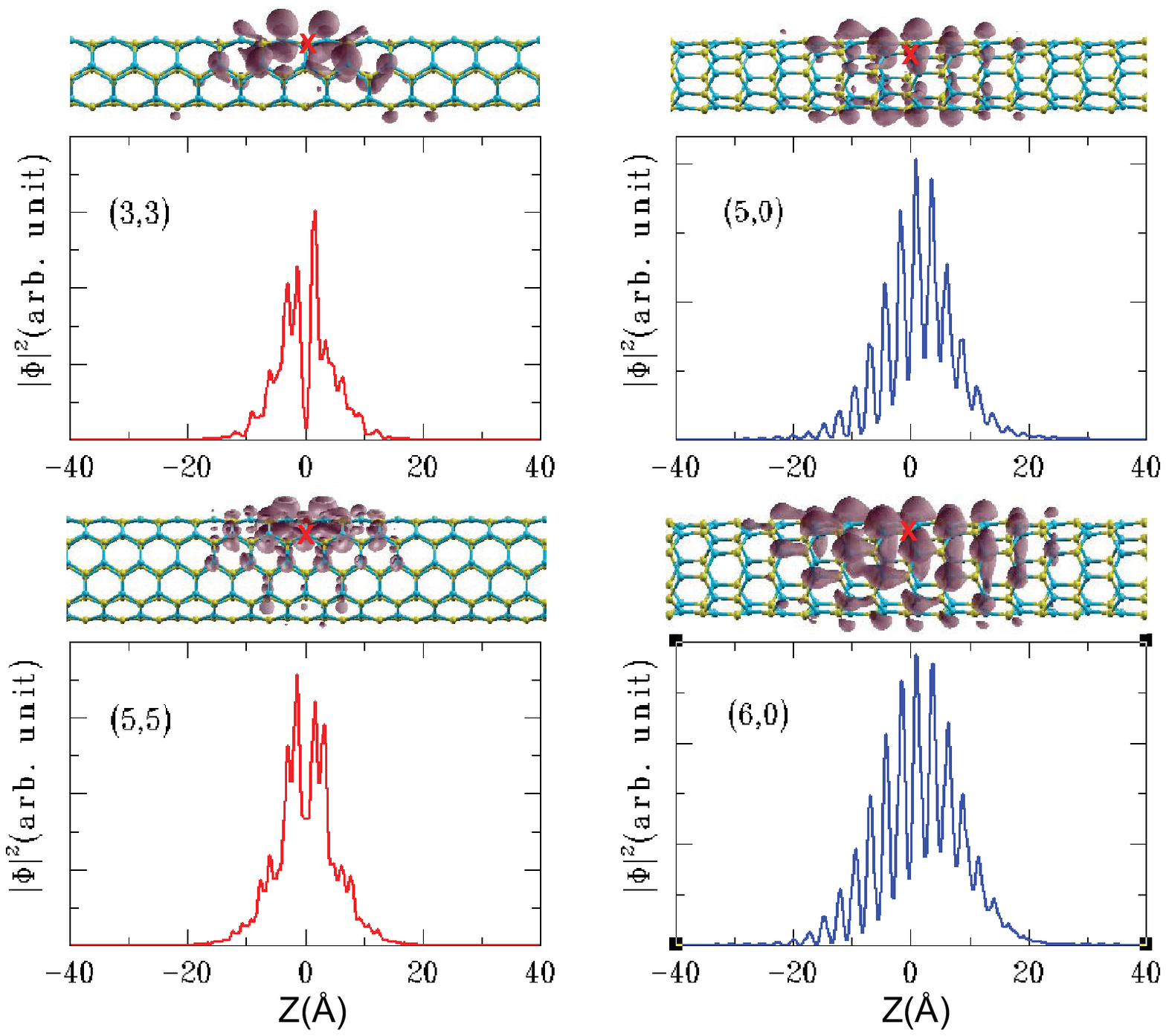}
\caption{(Color online) Isosurface plots of the electron distribution $|\Phi|^{2}$
of the $I_1$ exciton (see Fig. 7 and Fig. 8) with the fixed hole position
(indicated by a red cross) in the armchair (3,3) and (5,5) as well as
the zigzag (5,0) and (6,0) SiC-NTs. The corresponding integrated
intensities by averaging out the coordinates perpendicular to the tube axis,
are also shown. Here the hole position is set at 0. 
The Si and C atoms are represented by the cyan and yellow spheres, respectively.}
\end{figure}

\section{Conclusions}

In summary, we have employed the state-of-the art many-body GW and GW+BSE approaches to
study the quasiparticle band structure and optical properties, respectively, of 
the SiC sheet and related SiC nanotubes.
First of all, we find a direct quasiparticle band gap for the isolated 2D SiC sheet.
Our GW band structure calculations show that rather complicated orbital-dependent
self-energy corrections are needed in obtaining accurate quasiparticle
properties for the isolated SiC layer. 
The profile of the optical spectra is modified
dramatically when the electron-hole interaction is included.
In particular, a strongly bright bound exciton with a large binding energy 
(1.17 eV) is found to dominate the optical spectrum, because of the enhanced
overlap between the electron and hole orbitals due to reduced-dimensionality 
effects and also the existence of vacuum region which reduces the effective 
screening in the SiC sheet.

Secondly, 
the quasiparticle band structure of the small-radius armchair and zigzag SiC-NTs
is systematically studied within the GW approximation. 
Our detailed analysis on the charge density distributions reveals that a
curvate-induced orbital rehybridization plays a vital role on determining the
band gap of the small-radius SiC-NTs with different chiralities at both the 
LDA and GW levels. In particular, the quasiparticle band gaps as a function of 
tube diameter behave very differently, depending on the chirality of the
SiC-NTs concerned. 

Finally, the calculated photoexcited spectra are consisted of discrete exciton peaks,
thereby indicating strong excitonic effects in both the armchair and zigzag SiC-NTs.
The optical absorption spectra of the small radius armchair and zigzag SiC-NTs
are dominated by the first bright bound exciton $I_{1}$ with 
a significant binding energy up to $\sim$2.0 eV. The highly asymmetric
charge distribution of the exciton $I_{1}$ in the (3,3) SiC-NT is found to be
a consequence of the strong coupling of the four optically allowed inter-subband
transitions. Interestingly, we also find a quasi-zero dimensional bound character of 
the first bright exciton in the SiC-NTs and curvature-induced symmetry breaking 
affects the shape and size of this bound exciton. 
Moreover, our GW and BSE calculations demonstrate that the simple zone-folding approach 
fails in predicting the low energy exciton characters in very small zigzag SiC-NTs, mainly
because of the strong curvature-induced orbital rehybridization in these nanotubes.
We believe that the large excitation 
energy of $\sim 3.0$ eV of the first bright exciton, with no dark exciton below it, 
may make the small-radius armchair SiC-NTs candidates for optical devices working in the UV regime. 
In contrast, the numerous dark excitons below $I_{1}$ in the zigzag SiC-NTs 
may lead to potential applications in tunable optelectric devices ranging 
from infrared to UV frequencies by external perturbations. 

\begin{acknowledgments}
We thank Jack Deslippe for helpful discussions on performing the GW+BSE
calculations in BerkeleyGW code. H. C. H. and G. Y. G. thank the National Science Council
and NCTS of ROC for support, and also NCHC of ROC for CPU time. S.G.L. is supported by the Director, Office of Science, Office of Basic Energy Sciences, Materials Sciences and Engineering Division, U.S. Department of Energy under Contract No. DE-AC02-05CH11231. 
\end{acknowledgments}


\begin{references}
\bibitem{Hsueh} Here we have also performed the LDA, GW and BSE calculations for the bulk 2H SiC,
and find the LDA and GW band gaps to be 2.10 and 3.17 eV, respectively. The BSE calculation
predicts the exciton binding energy in 2H SiC to be 0.1 eV.
\end{references}

\end{document}